# Orthogonal Fault Tolerance for Dynamically Adaptive Systems

Sobia K Khan
School of Computing and Communications Systems, Lancaster University
Infolab21, Lancaster, LA1 4WA, U.K.
s.khan8@lancaster.ac.uk

*Abstract:* In dynamic systems that adapt to users' needs and changing environments, dependability needs cannot be avoided. This paper proposes an orthogonal fault tolerance model as a means to manage and reason about multiple fault tolerance mechanisms that co-exist in dynamically adaptive systems. One of the key challenges associated with dynamically evolving fault tolerance needs is the feature interaction problem arising from the integration of fault tolerance features. The proposed approach provides a separation of fault tolerance concerns to study the effects of integrated fault tolerance on the system. This approach uses state machine and operational semantics to reason about these interactions and inconsistencies. The proposed approach is supported by the tool NuSMV to simulate and verify the state machines against logic statements.

## I. INTRODUCTION

Recent years have seen a significant increase in dynamically adaptive systems – whereby the system's behavior is changed in response to its operational context, user requirements, or needs of other systems and services with which it interacts. This is particularly true of systems that operate in volatile and critical environments, such as those for crisis management, incident response or cyber security. Such systems must adapt in the presence of threats and faults and be able to react to hazardous situations. Given their safety and mission critical nature, one cannot ignore the need for dependability and trustworthiness in such systems. Dynamic adaptation introduces new interdependencies and interactions between features that lead to new threats and hazards. There is a need for a fault tolerance capability to handle those faults and failures dynamically at runtime. However, the very introduction of new fault tolerance mechanisms for this purpose leads to possible interdependencies and inconsistencies since many features, including already existing fault tolerance features, are influenced by the change. Consequently, such an adaptation of fault tolerance features can lead to new faults.

In this paper, we propose an orthogonal fault tolerance modeling (OFTM) approach to address the problems introduced by dynamically adapting fault tolerance features in a system. OFTM provides a separation of fault tolerance features from other features in the system. We propose an operational semantics and a composition mechanism to compose our new features with features of a running system. This separation of concerns ensures that the adaptation of fault tolerance features can be reasoned about independently before their incorporation into a running system.

The novel contributions of our approach are as follows:
- It is the first paper to propose orthogonal fault tolerance models as means to manage and reason about multiple fault tolerance mechanisms that may need to co-exist in a dynamically adaptive system.
- It provides a separation of concerns to study the effect of new or modified fault tolerance features on the system without adding complexity to the running system.
- It explicitly deals with feature interactions arising from incorporation of new/updated fault tolerance features into a dynamically adaptive system.

The rest of the paper is organized as follows. Section II presents a motivating example; section III discusses related work; section IV describes our proposed OFTM approach while section V concludes the paper and identifies directions for future work.

## II. MOTIVATING EXAMPLE

Our motivating example is that of a Home Automation scenario, where a house is fully equipped with a set of electrical sensors and actuators. Initially the *Lighting by Presence* feature is on (which senses whether the occupants are in the house or not before turning the lighting on or off) and, for security, the *Sensing* and *Silent Alarm* features are on. We assume that a Recovery Block fault tolerance technique is used for the Light Controller component. In such a technique, states are saved for check-pointing and backward recovery. We use state machines to depict the behavior of these features.

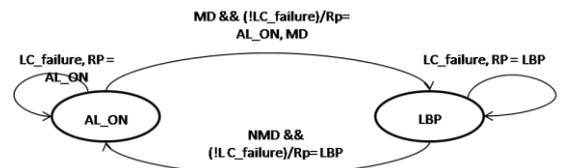

Fig 1: Light Controller (LC) State Machine

Key: LC: Light Controller, RP: Recovery Point, AL_ON: Automated Lights On, (N)MD: (No) Motion Detection, LBP: Lighting by Presence

The state machine in Fig. 1 shows the Light Controller component with Recovery Block fault tolerance. The Recovery Points (RP) for each state are saved as safe states, so that they may be recovered on component failure.

Now let us consider a dynamic adaptation scenario where a new feature, *Presence Simulation* (for security), is dynamically added. This feature simulates that the home is occupied by turning lights on and off while the occupants are away. This further requires the *Automated Lights* feature to be enabled. With the addition *Presence Simulation* feature, we add a further Recovery Block mechanism (see Fig. 2).

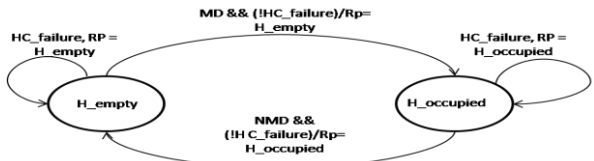

Fig 2: Home Controller (HC) State Machine

Key: HC: Home Controller, RP: Recovery Point, (N)MD: (No) Motion Detection, H: Home (empty/occupied)

When composed together, it is expected that, if the home is empty, then the *Automated Lights* feature will be enabled; otherwise the *Lighting by Presence* feature will be utilized. Both Light Controller and Home Status Controller components use the motion detection sensors to perform their required tasks. The Home Controller indicates whether the status of the home is empty or occupied. On the basis of this information, the Illumination and Security components carry out their tasks.

Fig. 3 shows the manual composition of these two state machines. If one of the controllers fails and recovers the previous safe state while the other is working properly, their interaction leads to a feature interaction problem.

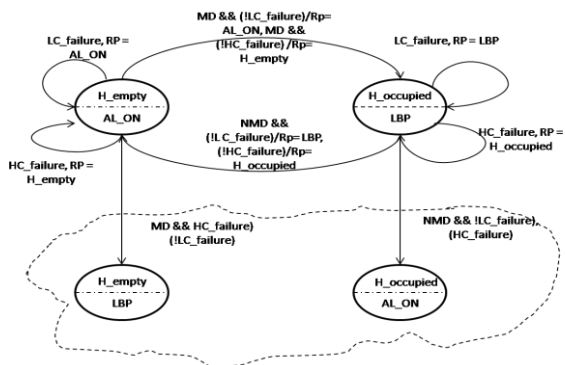

Fig 3: Composed State Machine

The states inside the dotted line are inconsistent states: (i) the home is empty but the *Lighting by Presence* feature is on; (ii) the home is occupied, but the *Automatic Lights* feature is on. These interactions arise not simply because of the dynamic adaptation itself, but due to the fault tolerance features introduced as a consequence of (or to support) the features added during that adaptation.

III. RELATED WORK

In [1], Kim and Lawrence describe the notion of adaptive fault tolerance, the contexts in which it is required and associated challenges. Since then a number of approaches have focused on this topic. The Simplex architecture [2] provides a reliable upgrading of a control system plus specific fault tolerance capability pertaining to timing faults and semantics faults. AFTM is a CORBA-based fault tolerance middleware [3] for dynamically adaptive fault tolerance in object-oriented real-time distributed applications. However, unlike our OFTM, both Simplex and AFTM an approach for adaptive fault tolerance in Object Oriented distributed systems are application-specific and do not address the issues of feature interactions arising from dynamic adaptation of fault-tolerance features.

Pareaud et al. [4] present a component based software engineering technique that relies on a reflective framework to provide fault tolerant software. In contrast, Ren et al. [5] provide an architecture approach, AQuA, for dependable object-oriented distributed systems. However, neither approach supports reasoning about the impact of fault tolerance adaptation on the overall architecture of the system. Other approaches such as those discussed in [6, 7, 8] lack the flexibility to provide fault tolerance capability in dynamic environments. Nor do they support reasoning about fault tolerance changes and analysis of potential feature interactions before these changes are introduced into the running system.

IV. PROPOSED APPROACH

A. *Orthogonal Fault Tolerance Model(OFTM)*

Our OFTM is inspired by Brito et al's feature model for fault tolerance mechanisms [15]. The OFTM captures the various features found in fault tolerance mechanisms reported in literature, as well as the relationships between features. As a result, not only known fault tolerance mechanisms but also bespoke ones can be derived from the OFTM. The ability to derive bespoke fault tolerance mechanisms is particularly pertinent to highly volatile situations such as crisis management scenarios where previously unknown configurations may need to be deployed and a suitable fault tolerance mechanism configured to suit the configuration in question [9].

To design our OFTM, we first listed various features of different fault tolerance mechanisms. We focused on software fault tolerance that uses design diversity to give system users continuous services. Design diversity mechanisms are mainly developed to deal with the design faults. By using design diversity, we can assure the coincident failure is rare in the presence of different software variants [16]. Our focus is on the design level rather than the implementation and testing level. Table 1 shows the main characteristics of some of the software fault tolerance mechanisms that we have considered so far.

TABLE 1: FAULT TOLERANCE MECHANISMS FEATURES

| | Execution Scheme | Error Processing Technique | Judgment Criteria | Checkpoints |
|---|---|---|---|---|
| Recovery Block | Sequential/ Parallel | Backward | AT/Voter | Yes |
| N-Version Programming | Sequential/ Parallel | Forward | AT/Voter | No |
| N-self checking Programming | Sequential/ Parallel | Forward/ Backward | AT/Voter/ Comparison | Yes/No |
| Distributed Recovery Block | Sequential | Forward | AT | No |
| Consensus Recovery Block | Parallel | Forward/ Backward | AT/Voter/ Comparison | Yes |

As shown above in Table: 1 there are four main features in the OFTM. The *Error Processing Technique* feature presents the different schemes for error processing like backward and forward error recovery along with acceptance test, voting and comparison. The *Execution Scheme* feature covers the two possible means for execution of fault tolerance behavior: sequential and parallel. The *Judgment Criteria* feature captures how acceptance tests should be performed either with absolute or relative criteria. The *Checkpoints* feature saves the states which can be recovered on failure.

We note that the OFTM itself can be inconsistent owing to multiple, potentially conflicting, fault tolerance mechanisms populating it, but any chosen fault tolerance configuration derived from the OFTM must be consistent. The OFTM enables reasoning about such interactions and inconsistency before composition takes place through the operational semantics associated with its constituent features.

## B. Operational Semantics, Composition and NuSMV

To show the structure and behavior of the components of dynamic system, we express them in terms of state machines; the fault tolerance mechanisms are also expressed in terms of state machines. The different features of fault tolerance are then composed with operational semantics, and verified against logic statements for correctness based on their relationship constraints. We are also using preprocessors to express the relationships and dependency rules between different features of fault tolerance mechanisms.

Based on our operational semantics, we can compose our system and fault tolerance state machines. Initially we refine the operational semantics into a Labeled Transition System (LTS), and then describe its translation to the input language of model checking tool called NuSMV [11, 12] – to simulate our state machines, perform the composition, and find any inconsistencies. We have chosen this tool because it deals with CTL (computation tree logic) as well as LTL (linear temporal language) [13, 14], allowing the analysis of interactions, deadlocks and other correctness properties.

The overall context of our proposed approach is a dynamic system in which both the main system and the fault tolerance mechanisms are expressed using state machines. Dynamic runtime adaptation (representing either changes to system components and/or fault tolerance needs) is modeled and then composed using operational semantics. The models are then verified for correctness, including the absence of inconsistencies and interactions, with the help of the model checking tool NuSMV.

## V. CONCLUSION AND FUTURE WORK

In this paper, we have proposed an orthogonal fault tolerance approach for dynamically adaptive systems. The main purpose of this method is to solve the problem of feature interaction arising from the incorporation of fault tolerance in dynamic systems at runtime. We have illustrated this problem with a simple case study of home automation. The proposed OFTM approach aims to enable reasoning about such interactions before composition takes place.

Future work involves completing the design of operational semantics for different combinations of fault tolerance mechanisms, and subsequently simulating and verifying the composed state machines in NuSMV for correctness. We also plan to extend the evaluation of our approach using a more realistically sized case study of home automation, and also in the context of home care systems.